\documentclass{aastex62}

\received{...}
\revised{...}
\accepted{...}

\submitjournal{ApJ}

\shorttitle{A Warm Flux Rope Eruption}
\shortauthors{Li et al.}

\begin{document}

\title{Eruption of a million-Kelvin warm magnetic flux rope on the Sun}

\correspondingauthor{Leping Li}
\email{lepingli@nao.cas.cn}

\author[0000-0001-5776-056X]{Leping Li}
\affil{National Astronomical Observatories, Chinese Academy of Sciences, Beijing 100101, Peoples's Republic of China}
\affiliation{Key Laboratory of Solar Activity and Space Weather, National Space Science Center, Chinese Academy of Sciences, Beijing 100190, People's Republic of China}
\affiliation{University of Chinese Academy of Sciences, Beijing 100049, Peoples's Republic of China}

\author[0000-0001-5705-661X]{Hongqiang Song}
\affiliation{Shandong Provincial Key Laboratory of Optical Astronomy and Solar-Terrestrial Environment, and Institute of Space Sciences, Shandong University, Weihai, Shandong 264209, Peoples's Republic of China}

\author[0000-0001-9921-0937]{Hardi Peter}
\affiliation{Max Planck Institute for Solar System Research, D-37077 G\"{o}ttingen, Germany}

\author[0000-0002-9270-6785]{Lakshmi Pradeep Chitta}
\affiliation{Max Planck Institute for Solar System Research, D-37077 G\"{o}ttingen, Germany}

\author[0000-0003-2837-7136]{Xin Cheng}
\affiliation{School of Astronomy and Space Science, Nanjing University, Nanjing 210093, Peoples's Republic of China}

\author[0000-0002-4230-2520]{Zhentong Li}
\affiliation{Key Laboratory of Dark Matter and Space Astronomy, Purple Mountain Observatory, Chinese Academy of Sciences, Nanjing 210023, People's Republic of China}
\affiliation{School of Astronomy and Space Science, University of Science and Technology of China, Hefei 230026, People's Republic of China}

\author[0000-0001-8228-565X]{Guiping Zhou}
\affil{National Astronomical Observatories, Chinese Academy of Sciences, Beijing 100101, Peoples's Republic of China}
\affiliation{Key Laboratory of Solar Activity and Space Weather, National Space Science Center, Chinese Academy of Sciences, Beijing 100190, People's Republic of China}
\affiliation{University of Chinese Academy of Sciences, Beijing 100049, Peoples's Republic of China}

\begin{abstract}

Solar magnetic flux rope (MFR) plays a central role in the physics of coronal mass ejections (CMEs).
It mainly includes a cold filament at typical chromospheric temperatures ($\sim$10000 K) and a hot channel at high coronal temperatures ($\sim$10\,MK). 
The warm MFR at quiescent coronal temperatures of a million Kelvin is, however,  rarely reported.
In this study, using multiwavelength images from Atmospheric Imaging Assembly (AIA) on board the Solar Dynamic Observatory (SDO) and Extreme Ultraviolet Imager (EUVI) on board the Solar Terrestrial Relations Observatory-A (STEREO-A), we present an eruption of a warm channel, that represents an MFR with quiescent coronal temperatures ($\sim$0.6-2.5\,MK).
On 2022 May 8, we observed the failed eruption of a hot channel, with the average temperature and emission measure (EM) of 10\,MK and 1.1$\times$10$^{28}$\,cm$^{-5}$, using AIA high-temperature images in active region (AR) 13007. 
This failed eruption was associated with a C8.2 flare, with no CME.
Subsequently, we observed a warm channel that appeared in AIA and EUVI low-temperature images, rather than AIA high-temperature images.
It then erupted, and transformed toward a semi-circular shape.
An associated C2.1 flare, along with the signatures of magnetic reconnection in AIA high-temperature images, were identified.
Additionally, we observed a CME associated with this event.
Compared with the hot channel, the warm channel is cooler and rarer with the average temperature and EM of 1.7 (1.6) MK and 2.0$\times$10$^{26}$ (2.3$\times$10$^{26}$) cm$^{-5}$.
All the results suggest an unambiguous observation of the million-Kelvin warm MFR, that erupted as a CME, and fill a gap in the temperature domain of coronal MFRs.

\end{abstract}


\section{Introduction} \label{sec:int}

Solar magnetic flux rope (MFR) is a coherent helical structure with magnetic field lines wound around a central axis \citep{1996ApJ...464L.199R, 2014ApJ...784L..36Y, 2016ApJ...818..148L}.
It is usually located upon the magnetic polarity inversion lines, and sometimes erupts outward explosively \citep{2015ApJ...808..117N, 2016NatPh..12..847L, 2016ApJ...829L..33L}.
The successful MFR eruption produces a coronal mass ejection (CME), that is the major driver of severe space weather \citep{2020SSRv..216..131P}.
The CME often reveals a classic three-part structure, including the bright front, dark cavity and bright core, in white-light coronagraphs \citep{2012A&A...539A...7L, 2013SoPh..284..179V}.
Among them, the dark cavity and bright core are generally interpreted as manifestations of the MFR \citep{2011LRSP....8....1C}. 
Recently, \citet{2019ApJ...883...43S, 2019ApJ...887..124S} suggested that the bright core corresponds solely to the MFR. 
In the physics of CMEs, the MFR thus plays a central role. 

In solar physics, the MFR mainly contains two activities, i.e., the filament (prominence) and hot channel \citep{2015A&A...580A...2Z, 2017ScChD..60.1383C, 2018ApJ...863..192L}.
The filament is composed of cooler, denser chromospheric material suspended in the hotter, rarer corona \citep[see reviews in][and references therein]{2010SSRv..151..333M, 2020RAA....20..166C}.
It is commonly thought to be supported in magnetic dips of the MFR \citep{2014ApJ...780..130X, 2018ApJ...856..179Z, 2019ApJ...883..104S, 2019ApJ...885L..11S}.
This is verified observationally by the filament rotation around its central axis during the eruption \citep{2013ApJ...773..162B, 2014ApJ...797...52Y, 2015ApJ...805....4Z, 2016NatCo...711837X}.
Moreover, tracing material from a surge and a failed filament eruption, \citet{2013ApJ...770L..25L} tracked out the helical structures of two MFRs that support the filaments.
Employing Atmospheric Imaging Assembly \citep[AIA;][]{2012SoPh..275...17L} images on board the Solar Dynamic Observatory \citep[SDO;][]{2012SoPh..275....3P}, \citet{2011ApJ...732L..25C} and \citet{2012NatCo...3..747Z} reported the hot channels (or blobs), and identified them as a type of MFR.
Using the differential emission measure (DEM) analysis, \citet{2012ApJ...761...62C} diagnosed the temperature and emission measure (EM) of the hot channels.
They found that the hot channel emission has a DEM-weighted average temperature larger than 8\,MK originating from a broad temperature range.
Recently, lots of hot channels have been detected and investigated in AIA high-temperature, e.g., 94 and 131\,\AA, images \citep{2014ApJ...784...48S, 2015ApJ...808..117N}.

Two types of MFR, i.e., the filament and hot channel, have the characteristic temperatures of 7500-9000\,K and $\textgreater$8\,MK, respectively, corresponding to the typical chromospheric and high coronal temperatures \citep{2014LRSP...11....1P, 2017ScChD..60.1383C}.
This introduces a question of whether a type of MFR, that has the characteristic quiescent coronal temperatures of a million Kelvin, between these two temperature ranges, exists.
The investigation of this type of MFR could be important because it may provide a more complete picture of the physics related to MFR.
Statistically searching for hot channels in 141 M-class and X-class flares, \citet{2015ApJ...808..117N} noticed that there are 24 eruptive events without the presence of a hot channel or a filament, and 20 out of the 24 events were associated with CMEs.
They proposed that the MFR exists during these CME/flare events but it is simply too cool to be detected in the 131\,\AA~passband and possibly too hot to show in the 304\,\AA~passband.
The detection of this type of MFR is hence essential for understanding the CME/flare events without eruptions of hot channels and/or filaments.

In a comprehensive radiative magnetohydrodynamic (MHD) simulation of flare-productive active regions (ARs), an eruption of the multithermal MFR took place in the corona \citep{2023ApJ...950L...3C}. 
This suggests that the MFR is a multithermal structure.
Using three-dimensional MHD simulations, \citet{2016ApJ...823...22X} studied the filament formation, and noticed that the MFR, that supports the filament, has temperatures of $\sim$1-2\,MK; see Figures\,3a and 5 in \citet{2016ApJ...823...22X}.
Investigating the eruption of MFRs underlying coronal streamers, \citet{2017ApJ...844...26F} found the formation of a twisted flux, that newly added to the MFR, with a temperature of $\sim$3.3\,MK; see Figure\,8 in \citet{2017ApJ...844...26F}.
Moreover, during the eruption, the MFR has a mean temperature of $\sim$2\,MK in the range of $\sim$1-4\,MK; see Figures\,11-13 in \citet{2017ApJ...844...26F}.
Numerical simulations suggest that the MFR, with the characteristic temperatures between those of the hot channel and filament, exists.

Using the SDO and Solar Terrestrial Relations Observatory \citep[STEREO;][]{2008SSRv..136....5K} A and B observations from three viewing angles, \citet{2013A&A...552L..11L} analyzed an eruption of two hot channels, and found both hot and cool components of the MFRs.
Analyzing a successful eruption, \citet{2014ApJ...794..149C} investigated the relationship between the hot channel and its associated filament, and suggested that the hot channel hosts the filament near its bottom.
Moreover, they detected the MFR component in AIA low-temperature, e.g., 335\,\AA, images.
Employing observations from two vantage perspectives, i.e., edge-on from the SDO and face-on from the STEREO-A, \citet{2022ApJ...933...68S} noticed that the erupting coronal cavity, hosting a filament at its bottom, is recorded as a channel-like structure in extreme ultraviolet (UV) (EUV), e.g., AIA 193\,\AA~and EUV Imager \citep[EUVI;][]{2008SSRv..136...67H} 195\,\AA, images.
Recently, \citet{2022ApJ...941L...1L} studied the failed eruption of a multithermal MFR with no associated filament, and proposed that both the hot and warm channels twist together, constituting the same MFR showing plasma with high (8.4\,MK) and quiescent (2.0\,MK) coronal temperatures. 
The warm component of MFR during the hot channel and/or filament eruptions has been previously reported \citep{2013A&A...552L..11L, 2014ApJ...794..149C, 2022ApJ...941L...1L, 2022ApJ...933...68S}, revealing the multithermal nature of MFR \citep{2023ApJ...950L...3C}.
The eruption of a warm MFR with quiescent coronal temperatures of $\sim$1-2\,MK with no associated filament and hot channel eruption is, however, rarely presented.

In this study, we report an unambiguous observation of the eruption of a million-Kelvin warm MFR and its associated flare and CME on 2022 May 8.
The observations, results, and summary and discussion are shown in Sections \ref{sec:obs}-\ref{sec:sum}, respectively.

\section{Observations}\label{sec:obs}

SDO/AIA is a set of normal-incidence imaging telescopes, acquiring the solar atmospheric images in 10 wavelength passbands.
Sun Earth Connection Coronal and Heliospheric Investigation (SECCHI) EUVI on board the STEREO provides solar images in four EUV, i.e., 171, 195, 284 and 304\,\AA, passbands.
Different passbands show plasma at different temperatures, e.g., 131\,\AA~peaks at $\sim$10\,MK (Fe\,{\sc xxi}) and $\sim$0.6\,MK (Fe\,{\sc viii}), 94\,\AA~peaks at $\sim$7.2\,MK (Fe\,{\sc xvii}), 335\,\AA~peaks at $\sim$2.5\,MK (Fe\,{\sc xvi}), 211\,\AA~peaks at $\sim$1.9\,MK (Fe\,{\sc xiv}), 193 and 195\,\AA~peaks at $\sim$1.5\,MK (Fe\,{\sc xii}), 171\,\AA~peaks at $\sim$0.9\,MK (Fe\,{\sc ix}), 304\,\AA~peaks at $\sim$0.05\,MK (He\,{\sc ii}), and 1600\,\AA~peaks at $\sim$0.1\,MK (C\,{\sc iv} and continuum).
Spatial sampling of the AIA and EUVI images are separately 0.6\arcsec/pixel and 1.6\arcsec/pixel.
Time cadences of the AIA EUV and UV images and the EUVI 195\,\AA~images are 12 and 24\,s and 2.5 minutes, respectively, while those of the EUVI 171, 284 and 304\,\AA~images are larger and nonuniform.
In this study, we mainly employ the AIA 131, 94, 335, 211, 193, 171, 304 and 1600\,\AA, and the EUVI-A 195 and 304\,\AA~images to investigate the evolution of MFR eruption and its associated flare.
For AIA, level 1 data is downloaded, and then converted to level 1.5 data by using ``aia\_prep.pro" in the Solar Software (SSW).
To better illustrate the evolution, AIA and EUVI-A images are enhanced using the Multiscale Gaussian Normalization (MGN) technique \citep{2014SoPh..289.2945M}. 

GOES-16 1-8\,\AA~soft X-ray flux with time cadence of 1\,s is employed to study the flare associated with the MFR eruption. 
White-light coronagraphs from Large Angle and Spectrometric Coronagraph \citep[LASCO;][]{1995SoPh..162..357B} on board the Solar and Heliospheric Observatory \citep[SOHO;][]{1995SSRv...72...81D} and SECCHI COR1 and COR2 on board the STEREO-A are used to investigate the CME associated with the MFR eruption.
Here, we mainly employ the SOHO/LASCO C2 and STEREO-A/SECCHI COR2 white-light coronagraphs, with spatial sampling of 11.9\arcsec/pixel and 14.7\arcsec/pixel, and time cadences of $\sim$12 and $\sim$15 minutes, respectively.

\section{Results}\label{sec:res}

The SDO and STEREO-A satellites generally observe the Sun from different vantage points.
Figure\,\ref{f:general_information}(a) shows the positions of the two satellites at 22:00 UT on 2022 May 8.
The investigated MFR; see Figure\,\ref{f:general_information}(b), is located out of the southeastern solar limb in the FOV of SDO; see the red diamond in Figure\,\ref{f:general_information}(a).
Therefore, it is also observed by the STEREO-A; see Figure\,\ref{f:general_information}(a), at the heliographic position S19 E42 on the solar disk; see Figure\,\ref{f:general_information}(c).
Tracking the SDO and STEREO-A observations in the following days, we notice that the MFR is located in NOAA AR 13007; see Figures\,\ref{f:general_information}(b)-(c).
From 19:00 UT on May 8, two C-class flares are recorded succesively by the GOES-16 1-8\,\AA~soft X-ray flux; see Figure\,\ref{f:general_information}(d).
The first one is a C8.2 flare that started from 19:23 UT, peaked at 19:38\,UT; see the blue vertical dotted line in Figure\,\ref{f:general_information}(d), and ended at 19:52\,UT, lasting for 29 minutes.
The second one is a C2.1 flare that started from 21:40\,UT, peaked at 22:41\,UT; see the red vertical dotted line in Figure\,\ref{f:general_information}(d), and ended at 00:08\,UT on May 9, lasting for $\sim$2.5\,hr, much longer than that of the first flare. 

\subsection{Failed eruption of a hot channel}\label{sec:hce}

From $\sim$19:00 UT on May 8, a hot channel appeared in AIA high-temperature, i.e., 131 and 94\,\AA; see Figures\,\ref{f:hot_channel}(a)-(b), rather than low-temperature, e.g., 193\,\AA; see Figure\,\ref{f:hot_channel}(c), images out of the southeastern solar limb from AR 13007, with an elongated, twisted configuration; see Figures\,\ref{f:hot_channel}(d)-(e).
It then erupted toward the southeast.
Expanding loops are detected ahead of the erupting hot channel in AIA low-temperature, e.g., 193\,\AA, images; see Figure\,\ref{f:hot_channel}(f).
Along the AB direction in the red rectangle in Figure\,\ref{f:hot_channel}(a), a time-slice of AIA 131\,\AA~images is made, and displayed in Figure\,\ref{f:measurements}(a).
The hot channel first rose slowly with a mean speed of $\sim$28 km\,s$^{-1}$, and then accelerated to $\sim$152\,km\,s$^{-1}$ with a mean acceleration of $\sim$221\,m\,s$^{-2}$; see the red dotted line in Figure\,\ref{f:measurements}(a).
Subsequently, it decelerated gradually with a mean acceleration of -20\,m\,s$^{-2}$, and finally stopped and disappeared, suggesting a failed eruption; see the online animated version of Figure\,\ref{f:hot_channel}.
During the eruption, several fine structures of the hot channel are identified, marked by the red solid arrows in Figure\,\ref{f:measurements}(a).
Moreover, during the deceleration phase of the hot channel eruption, the expanding loops in AIA lower-temperature images also decelerated and stopped; see the online animated version of Figure\,\ref{f:warm_channel}. 
This also supports the failed eruption of the hot channel.

Using six AIA passbands, including 131, 94, 335, 211, 193 and 171\,\AA, we analyze the temperature and EM of the hot channel.
Here, we employ the DEM analysis using ``xrt\_dem\_iterative2.pro" \citep{2004IAUS..223..321W, 2012ApJ...761...62C}, 
that has been widely applied to temperature diagnostics of features such as the hot channels \citep{2012ApJ...761...62C, 2014ApJ...784...48S, 2022ApJ...941L...1L}, flares \citep{2014ApJ...786...73S}, and jets \citep{2014A&A...567A..11Z}.
The temperature range in our computations is 5.5$\le\log T (K)\le$7.5, which is mainly determined by the temperature response curves of the six AIA passbands \citep{2012ApJ...761...62C}.
The uncertainties in the DEM results mainly arise from the uncertainties of the temperature response functions of the AIA EUV passbands, e.g., the non-ionization equilibrium effects, as well as the determination of the foreground and background \citep[see more details in][and references therein]{2012ApJ...761...62C}.
Here, version 10 of the temperature response functions, calculated using CHIANTI v9.3, is employed.
In order to estimate the DEM uncertainties, 100 Monte Carlo realizations of the data are computed.
The DEM-weighted average temperature and total EM are separately calculated by using 
\begin{equation}
T_{average}=\frac{\int_{T_{min}}^{T_{max}}DEM(T)T\,dT}{\int_{T_{min}}^{T_{max}}DEM(T)\,dT} 
\end{equation}
and
\begin{equation}
EM=\int_{T_{min}}^{T_{max}}DEM(T)\,dT, 
\end{equation} 
where T$_{min}$ and T$_{max}$ are the lower and upper limits of the integration.
Please refer to \citet{2012ApJ...761...62C} for details.
In the following text, the temperature and the EM refer to the DEM-weighted average temperature and the total EM, respectively.

The hot channel region, enclosed by the blue rectangle in Figure\,\ref{f:hot_channel}(d), is chosen to calculate the DEM.
The nearby quiet-Sun region at the similar height from the solar limb to the selected hot channel region, enclosed by the pink rectangle in Figure\,\ref{f:hot_channel}(d), is chosen for computing the background emission, that is subtracted from the hot channel region.
In each region, the digital number counts in each of the six AIA passbands are normalized temporally by the exposure time and averaged spatially over all pixels.
In both regions, the signal for the six AIA passbands is above the noise levels.
Figure\,\ref{f:measurements}(c) displays the DEM curve of the hot channel region,
that is well constrained with small errors.
A double-peak distribution in the DEM curve appeared. 
The main peak lies at 10\,MK (6.6$\le\log T (K)\le$7.3), and the secondary peak, two orders of magnitude smaller than the main peak, appears at $\sim$0.6\,MK (5.5$\le\log T (K)\le$6.3).
Between them, the high temperature plasma corresponds to the hot channel, and the low temperature plasma corresponds to the ambient warm coronal loops that are superimposed on the hot channel along the line of sight (LOS); see Figures\,\ref{f:hot_channel}(c) and (f).
Over the temperature range of 6.6$\le\log T (K)\le$7.3, we calculate the  temperature and EM using equations (1) and (2).
The temperature is 10.0$\pm$0.1\,MK, consistent with the imaging observations that the hot channel appears only in the AIA high-temperature passbands. 
The EM of the hot channel region is (1.1$\pm$0.1)$\times$10$^{28}$\,cm$^{-5}$.
Here, the errors of the temperature and EM are calculated from the 100 Monte Carlo solutions.
We also compute the temperature and EM over the temperature range of 5.5$\le\log T (K)\le$7.5, and obtain the similar values of 9.8$\pm$0.2\,MK and (0.9$\pm$0.1)$\times$10$^{28}$\,cm$^{-5}$, respectively.
This indicates that hot plasma dominates the emission in the hot channel region.

Employing the EM, electron number density (n$_{e}$) of the hot channel is estimated using $n_{e}$=$\sqrt{\frac{EM}{D}}$, where $D$ is the LOS depth of the hot channel.
Assuming that the depth ($D$) equals the width ($W$) of the hot channel, then the density is $n_{e}$=$\sqrt{\frac{EM}{W}}$.
We measure the hot channel width at the place and time where and when we calculate the DEM.
First we get the intensity profile in the 131\,\AA~passband perpendicular to the hot channel.
Using the mean intensity surrounding the hot channel as the background emission, we subtract it from the intensity profile.
As the intensity profile of the selected hot channel region is affected by the surrounding structures, e.g., sub-structures of the hot channel, for simplicity, objectivity and accuracy, we fit the residual intensity profile using a single Gaussian, and obtain the FWHM of the single Gaussian fit as the width.
The cross-sectional width of the hot channel is measured to be 10.8$\pm$0.3\,Mm.
Here, the error of the width is estimated from the errors between the single Gaussian fit and the residual intensity profile.
Employ EM=(1.1$\pm$0.1)$\times$10$^{28}$\,cm$^{-5}$, the density is then estimated to be (3.2$\pm$0.2)$\times$10$^{9}$\,cm$^{-3}$.

The C8.2 flare; see Figure\,\ref{f:general_information}(d), is associated with the failed eruption of the hot channel.
Post-flare loops of the C8.2 flare appeared in AIA EUV images; see Figures\,\ref{f:hot_channel}(d)-(f).
Two flare ribbons formed at the endpoints of post-flare loops in AIA 1600\,\AA~images; see Figure\,\ref{f:hot_channel}(g).
No CME associated with the hot channel eruption is detected in LASCO white-light coronagraphs, further supporting that the hot channel eruption failed.
The flare-related activity is also observed by STEREO-A EUVI images; see the online animated version of Figure\,\ref{f:warm_channel}.
Although the hot channel in itself is not detected, the loop expansion, caused by the hot channel eruption, is identified in EUVI-A 195\,\AA~images; see Figure\,\ref{f:hot_channel}(i).
Moreover, the post-flare loops and flare ribbons associated with the hot channel eruption are also observed in EUVI-A 195 and 304\,\AA~images; see Figures\,\ref{f:hot_channel}(h)-(i).
No CME, associated with the hot channel eruption, is detected in STEREO-A COR1 and COR2 white-light coronagraphs.
The information of the hot channel eruption is listed in Table\,\ref{tab:eruptions}.

\subsection{Eruption of a warm channel}\label{sec:wce}

Shortly after the failed hot channel eruption, from $\sim$20:50 UT on May 8, a warm channel with fine structures appeared in AIA low-temperature, e.g., 211, 193 and 171\,\AA, rather than high-temperature, e.g., 94 and 131\,\AA, images; see Figures\,\ref{f:warm_channel}(a)-(c).
It shows plasma at quiescent, rather than high, coronal temperatures, identical to that previously reported in \citet{2022ApJ...941L...1L}.
The warm channel then erupted toward the southeast.
During the eruption, it transformed toward a semi-circular shape; see Figures\,\ref{f:warm_channel}(d)-(f).
Ahead of the eruption, compression front formed in AIA low-temperature images; see the online animated version of Figure\,\ref{f:warm_channel}.
Along the direction CD in the red rectangle in Figure\,\ref{f:warm_channel}(b), a time-slice of AIA 193\,\AA~images is made, and displayed in Figure\,\ref{f:measurements}(b), where the green dotted line outlines the eruption.
The warm channel first rose slowly with a mean speed of $\sim$25\,km\,s$^{-1}$, and then accelerated to $\sim$206\,km\,s$^{-1}$ with a mean acceleration of $\sim$44\,m\,s$^{-2}$.
Fine structures of the warm channel, denoted by the green solid arrows in Figure\,\ref{f:measurements}(b), are identified.
The compression front, running ahead of the erupting warm channel, is marked by the blue solid arrows in the blue rectangle, where the signal is enhanced for better illustration, in Figure\,\ref{f:measurements}(b).
It rose slowly with a mean speed of $\sim$17\,km\,s$^{-1}$, and then accelerated to $\sim$76\,km\,s$^{-1}$ with a mean acceleration of $\sim$16\,m\,s$^{-2}$.
The warm channel eruption and its associated compression front are also observed in EUVI-A 195\,\AA~images; see Figures\,\ref{f:warm_channel}(g)-(h) and the online animated version of Figure\,\ref{f:warm_channel}.
The information of the warm channel eruption is listed in Table\,\ref{tab:eruptions}.

In the blue rectangle in Figure\,\ref{f:warm_channel}(e), we calculate the DEM of the warm channel region, and show it in Figure\,\ref{f:measurements}(d).
Here, the background emission, that is subtracted from the warm channel region, is computed in the nearby quiet-Sun region at the similar height from the limb to the selected warm channel region; see the pink rectangle in Figure\,\ref{f:warm_channel}(e).
In both regions, the signal in the AIA 211, 193 and 171\,\AA~passbands is much higher than the noise levels, and that in the AIA 131, 94 and 335\,\AA~passbands is close to the noise levels.
Comparing with the DEM of the hot channel in Figure\,\ref{f:measurements}(c), the warm channel emission is well constrained in the temperature range of 5.8$\le\log T (K)\le$6.4 (0.6-2.5\,MK), rather than the low and high temperatures, due to the absence of signal in the AIA low and high temperature passbands, i.e., 131, 94 and 335\,\AA. 
Over the temperature range of 5.8$\le\log T (K)\le$6.4, we calculate the temperature using equation (1), and obtain a value of 1.7$\pm$0.1\,MK, consistent with the AIA and EUVI imaging observations that the warm channel is observable only in AIA and EUVI low-temperature, rather than AIA high-temperature, images.
The EM of the warm channel is (2.0$\pm$0.3)$\times$10$^{26}$\,cm$^{-5}$, much smaller than that of the hot channel.

As the DEM curve is poorly contrained in the low and high temperatures; see Figure\,\ref{f:measurements}(d), we employ an improved sparse inversion code \citep{2015ApJ...807..143C, 2018ApJ...856L..17S} to recalculate the DEM.
In this method, well-constrained DEM solutions are obtained in the case of using the new basis functions and tolerance control.
Please refer to \citet{2018ApJ...856L..17S} for details.
The DEM is shown as purple curve in Figure\,\ref{f:measurements}(d), where the pink rectangles represent the Monte Carlo solutions.
Over the temperature range of 5.8$\le\log T (K)\le$6.4, we calculate the temperature and EM using equations (1) and (2), and get values of 1.6 MK and 2.3$\times$10$^{26}$\,cm$^{-5}$, similar to those by using ``xrt\_dem\_iterative2.pro". 
Both temperatures (1.7 and 1.6\,MK) support that the warm channel shows plasma at quiescent coronal temperatures of $\sim$1-2\,MK.
Same as in Section\,\ref{sec:hce}, we measure the warm channel width at the time and place when and where we calculate the DEM in AIA 193\,\AA~images, and obtain a value of 29.3$\pm$1.4\,Mm.
Using EM=(2.0$\pm$0.3)$\times$10$^{26}$ (2.3$\times$10$^{26}$) cm$^{-5}$ and W=29.3$\pm$1.4\,Mm, we estimate the density of the warm channel, and get a value of (2.6$\pm$0.3)$\times$10$^{8}$ ((2.8$\pm$0.1)$\times$10$^{8}$) cm$^{-3}$, one order of magnitude smaller than that of the hot channel.
Furthermore, we employ the impoved sparse inversion code to recalculate the DEM of the hot channel region, and obtain the similar result to that in Section\,\ref{sec:hce}.

The C2.1 flare; see Figure\,\ref{f:general_information}(d), is associated with the warm channel eruption.
Two flare ribbons of the C2.1 flare are recorded in AIA 1600\,\AA~images; see Figure\,\ref{f:flare_cme}(a).
Post-flare loops, connecting two flare ribbons, appeared in AIA EUV images; see Figure\,\ref{f:flare_cme}(b).
Both of them are also detected in EUVI-A 304 and 195\,\AA~images; see Figures\,\ref{f:flare_cme}(c)-(d).
Comparing with the previous C8.2 flare associated with the failed hot channel eruption, this one is much extensive. 
The eastern ribbons of two flares are located at similar positions, however, the western ribbon of the C2.1 flare is located to the west of the western ribbon of the C8.2 flare; see Figures\,\ref{f:hot_channel}(d)-(i) and \ref{f:flare_cme}(a)-(d).
Moreover, the post-flare loops of the C2.1 flare is much longer and higher than those of the C8.2 flare; see Figures\,\ref{f:hot_channel} and \ref{f:flare_cme}(a)-(d).
Different from the failed hot channel eruption, a CME, associated with the successful warm channel eruption, is detected in LASCO C2 (C3) and STEREO-A COR2 (COR1) white-light coronagraphs; see Figures\,\ref{f:flare_cme}(e)-(f), with mean speeds of 482 and 593 km/s, respectively.
The speeds and shapes of the CME are slightly different observed from these two instruments; see the online animated version of Figure\,\ref{f:flare_cme}, due to the different viewing angles, time cadences and spatial sampling.

Different from the previous C8.2 flare, magnetic reconnection region was identified above the post-flare loops during the C2.1 flare in AIA high-temperature, i.e., 131 and 94\,\AA, images; see the online animated version of Figure\,\ref{f:reconnection}.
After the warm channel eruption, brightening took place at the top of post-flare loops that connect the flare ribbons; see Figures\,\ref{f:reconnection}(a1)-(b1), from $\sim$21:40\,UT, identical to the start time of the flare.
Bright loop-like structures then appeared successively from low to high above the post-flare loops; denoted by the red solid arrow in Figure\,\ref{f:reconnection}(a1).
They may represent the newly-reconnected loops, i.e., the newly-formed post-flare loops, during the flare.
In the blue rectangle in Figure\,\ref{f:reconnection}(a1), we calculate the DEM for the loop region, and display it in Figure\,\ref{f:reconnection_evolution}(a).
Here, the background emission, that is subtracted from the loop region, is computed in the nearby quiet-Sun region at the similar height from the limb to the selected loop region; see the pink rectangle in Figure\,\ref{f:reconnection}(a1).
For the following DEM calculations of the reconnection regions; see the blue rectangles in Figures\,\ref{f:reconnection}(a2)-(a4) and (b4), the same quiet-Sun region; see the pink rectangles in Figures\,\ref{f:reconnection}(a2)-(a4) and (b4), is chosen for computing the background emission.
For all the selected reconnection and background emission regions; see the blue and pink rectangles in Figure\,\ref{f:reconnection}, the signal for the six AIA passbands is above the noise levels.
The DEM curve of the loop region is well constrained in the whole temperature range, i.e., 5.5$\le\log T (K)\le$7.5.
Over this temperature range, we calculate the temperature using equation (1), and obtain a value of 11.8$\pm$1.2\,MK.
This indicates that the bright loop emission is mainly from plasma at high temperatures, consistent with the AIA imaging observations that the bright loop appears only in AIA high-temperature, rather than low-temperature, images.
The EM of the bright loop is (9.0$\pm$1.3)$\times$10$^{27}$\,cm$^{-5}$.
Same as in Section\,\ref{sec:hce}, we measure the width of the bright loop in AIA 131\,\AA~images at the place and time where and when we calculate the DEM, and obtain a value of 4.9$\pm$0.4\,Mm.
Employing EM=(9.0$\pm$1.3)$\times$10$^{27}$\,cm$^{-5}$ and W=4.9$\pm$0.4\,Mm, we estimate the density of the bright loop, and get a value of (4.3$\pm$0.5)$\times$10$^{9}$\,cm$^{-3}$.
The information of the reconnection region is listed in Table\,\ref{tab:reconnection}.

Along with the formation of bright loops, a cusp-shaped structure took place in AIA 131\,\AA~images; see Figures\,\ref{f:reconnection}(a2)-(c3), above the post-flare loops that appeared in AIA 171\,\AA~($\sim$0.9\,MK) images from $\sim$22:35\,UT.
In the blue rectangles in Figures\,\ref{f:reconnection}(a2)-(a3), we calculate the DEM for the cusp structure regions, and show them in Figures\,\ref{f:reconnection_evolution}(b)-(c).
Both DEM curves are well constrained in the whole temperature range.
Over the temperature range of 5.5$\le\log T (K)\le$7.5, we calculate the temperature and EM using equations (1) and (2).
The temperatures of the cusp structure are 9.5$\pm$0.2 and 8.5$\pm$0.2\,MK, and the EM are (3.5$\pm$0.2)$\times$10$^{28}$ and (2.0$\pm$0.2)$\times$10$^{28}$\,cm$^{-5}$, respectively.
Same as in Section\,\ref{sec:hce}, we measure the widths of the cusp structure in AIA 131\,\AA~images at the time and place when and where the DEM are calculated, and get respective values of 10.9$\pm$0.1 and 15.6$\pm$0.5\,Mm.
The densities of the cusp structure are then estimated separately to be (5.7$\pm$0.2)$\times$10$^{9}$ and (3.6$\pm$0.2)$\times$10$^{9}$\,cm$^{-3}$, similar to that of the bright loops. 

Subsequently, supra-arcade spikes appeared above the cusp structure; denoted by the red solid arrows in Figures\,\ref{f:reconnection}(a4)-(b4).
They may represent the reconnection region during the flare.
Along the supra-arcade spikes, material fell toward the solar surface; see the online animated version of Figure\,\ref{f:reconnection}.
In the blue rectangle in Figure\,\ref{f:reconnection}(a4), we calculate the DEM for the supra-arcade spike region, and show it in Figure\,\ref{f:reconnection_evolution}(d).
The DEM curve is well constrained in the temperature range of 5.5$\le\log T (K)\le$7.1.
Over this temperature range, we calculate the temperature and EM of the supra-arcade spikes using equations (1) and (2), and get values of
7.5$\pm$0.5\,MK and (5.0$\pm$1.5)$\times$10$^{27}$\,cm$^{-5}$, respectively.
Same as in Section\,\ref{sec:hce}, we measure the width of the supra-arcade spike in AIA 131\,\AA~images at the time and place when and where we calculate the DEM, and obtain a value of 12.6$\pm$3.5\,Mm.
Using EM=(5.0$\pm$1.5)$\times$10$^{27}$\,cm$^{-5}$ and W=12.6$\pm$3.5\,Mm, we estimate the density of the supra-arcade spike, and get a value of (2.1$\pm$0.6)$\times$10$^{9}$\,cm$^{-3}$, a little smaller than those of the bright loops and cusp structures; see Table\,\ref{tab:reconnection}.

In the blue rectangle in Figure\,\ref{f:reconnection}(b4), we calculate the DEM for the supra-arcade spike region, and obtain the temperature and EM using equations (1) and (2).
Here, the temperature range of 5.6$\le\log T (K)\le$7.4 is used for the convenience of calculations.
The temporal evolution of the temperature and EM is displayed in Figure\,\ref{f:reconnection_evolution}(e) as red diamonds and blue squares, respectively.
We overlay the GOES-16 1-8\,\AA~soft X-ray flux in Figure\,\ref{f:reconnection_evolution}(e) as a black curve.
Before the increase, the temperature and EM are almost constant with values of 1.8\,MK and 1.0$\times$10$^{27}$\,cm$^{-5}$, which are the characteristic parameters of the quiet-Sun region before the appearance of the reconnection region.
They started to increase from 22:09\,UT, later than the start time (21:40\,UT) of the C2.1 flare, with values of 7.9\,MK and 3.1$\times$10$^{27}$\,cm$^{-5}$. 
This delay is due to the fact that the chosen region for computing the DEM  is a small reconnection region above the post-flare loops, rather than the whole flare region.
The temperature then quickly increased to the peak of 11\,MK at 22:15 UT, earlier than the peak time (22:41\,UT) of the C2.1 flare, in 6 minutes.
After the peak, it decreased slowly to 5.7\,MK in $\sim$2\,hr.
The EM increased more slowly than the temperature, and peaked at 22:49\,UT, later than the peak times of the temperature (22:15\,UT) and the C2.1 flare (22:41\,UT), with a peak value of 2.0$\times$10$^{28}$\,cm$^{-5}$.
It slowly decreased after the peak to 1.2$\times$10$^{28}$\,cm$^{-5}$ in $\sim$1\,hr, and  increased again slightly after $\sim$23:40\,UT. 
The reincrease of the EM during the decay phase of the C2.1 flare may be caused by the rising of post-flare loops into the region that was chosen for calculating the DEM.
%

\section{Summary and discussion}\label{sec:sum}

Employing SDO/AIA and STEREO-A/EUVI images, and SOHO/LASCO and STEREO-A/COR2 white-light coronagraphs, we investigate the eruption of the warm channel, showing plasma at quiescent coronal temperatures of $\sim$0.6-2.5\,MK, in AR 13007, and its associated flare and CME.
On 2022 May 8, the hot channel, with the temperature and EM of 10.0$\pm$0.1\,MK and (1.1$\pm$0.1)$\times$10$^{28}$\,cm$^{-5}$, appeared in high-temperature, i.e., 94 and 131\,\AA, rather than low-temperature, e.g., 211, 193 (195) and 171\,\AA, images.
It slowly rose, quickly erupted, and finally stopped and disappeared, indicating the failed eruption of the hot channel.
The C8.2 flare and no CME are associated with the failed eruption.
Subsequently, the warm channel, with the temperature and EM of 1.7$\pm$0.1 (1.6) MK and (2.0$\pm$0.3)$\times$10$^{26}$ (2.3$\times$10$^{26}$) cm$^{-5}$, appeared in low-temperature, e.g., 211, 193 (195) and 171\,\AA, rather than high-temperature, i.e., 94 and 131\,\AA, images.
It rose slowly, and then erupted quickly, with the formation of the compression front running ahead of the warm channel in low-temperature images.
Associated with the successful eruption of the warm channel, the C2.1 flare and CME are identified.
Different from the previous C8.2 flare, the reconnection region above the post-flare loops of the C2.1 flare, including the newly-reconnected loops, the cusp structure, and the supra-arcade spikes, is detected.
The temperature and EM of the reconnection region, and their temporal evolution are then analyzed.

A million-Kelvin warm MFR is reported.
In this study, the property of the hot channel indicates an MFR with high coronal temperatures (10.0$\pm$0.1\,MK); see Section\,\ref{sec:hce}, consistent with those previously investigated \citep{2012ApJ...761...62C, 2015ApJ...808..117N, 2022ApJ...941L...1L}.  
During the eruption of the warm channel, no associated eruption of the filament and hot channel is identified; see Section\,\ref{sec:wce}.
This shows an unambiguous observation of the warm MFR with quiescent coronal temperatures, filling a gap in the temperature domain of MFR.
But the frequency of occurrence of warm MFRs in comparison with hot MFRs needs to be further explored.
The temperature and density of the warm channel are 1.7$\pm$0.1 (1.6) MK and (2.6$\pm$0.3)$\times$10$^{8}$ ((2.8$\pm$0.1)$\times$10$^{8}$) cm$^{-3}$, much smaller than those (10.0$\pm$0.1\,MK and (3.2$\pm$0.2)$\times$10$^{9}$\,cm$^{-3}$) of the hot channel; see Table\,\ref{tab:eruptions}.
Here, the heights from the limb, where we measured the densities of the hot ($\sim$50\,Mm) and warm ($\sim$180\,Mm) channels, should be considered, because the densities of the hot and warm channels, that are expanding features during the eruptions, vary with heights.
Comparing with the warm component of the multithermal MFR reported previously in \citet{2022ApJ...941L...1L}, the warm MFR here has a similar temperature, but a smaller density.
To further understand the nature of warm MFR with quiescent coronal temperatures, a statistical work of the warm MFR is needed in the future.
Investigating the frequency of hot MFRs in larger flares, \citet{2015ApJ...808..117N} found that 24 eruptive events, most of which are associated with CMEs, have no associated hot channel and filament eruption.
They suggested that the warm MFR exists in these events, with temperatures between those of the hot channel and filament. 
In this study, the observation of the 0.6-2.5\,MK warm MFR can be employed to explain the eruptive flares or CMEs without associated eruption of hot channel and filament.
Furthermore, to understand the reason why some MFRs are heated to the high coronal temperatures of $\sim$10\,MK as the hot channels, and why some MFRs are heated to the quiescent coronal temperatures of $\sim$1-2\,MK as the warm channels, more observations and numerical simulations of the MFR are needed in the future.

The sympathetic eruptions of the hot and warm channels are observed.
In this study, both the hot and warm channels underwent the slow rise phase and then the main acceleration phase during the eruptions; see Figures\,\ref{f:measurements}(a)-(b), consistent with those in \citet{2012NatCo...3..747Z} and \citet{2020ApJ...894...85C}.
They, however, show the failed and successful eruptions, respectively.
This may be due to the smaller erupting speed of the hot channel compared to that of the warm channel; see Table\,\ref{tab:eruptions}.
The erupting speed may be hence crucial for whether the eruption is a successful or failed eruption. 
The observations suggest that the failed eruption of the hot channel triggers the subsequent successful eruption of the warm channel.
The eastern ribbons of the C8.2 and C2.1 flares, separately associated with the hot and warm channel eruptions, are located at the similar positions, however, the western ribbon of the C2.1 flare is located to the west of that of the C8.2 flare; see Sections\,\ref{sec:hce} and \ref{sec:wce}.
This indicates that the eastern endpoint of the overlying field of the hot and warm channels roots in the similar magnetic field, where the eastern ribbons of two flares are located.
The western endpoint of the overlying field of the warm channel roots in the magnetic field, that is located to the west of the magnetic field, where the western endpoint of the overlying field of the hot channel roots; see the western ribbons of two flares.
The overlying field of the warm channel thus contains that of the hot channel, and both of them have one similar endpoint, i.e., the eastern endpoint.
The stronger overlying field of the hot channel, compared to that of the warm channel, may result in the failed eruption of the hot channel.
The overlying field thus may be also crucial for whether the eruption is a successful or failed eruption.
The failed hot channel eruption pushed the overlying field of the hot channel, and also the warm channel, upward and/or sideward.
This changed the equilibrium state of the warm channel, and caused the warm channel to erupt, showing the sympathetic eruptions \citep{2009RAA.....9..603J, 2011ApJ...739L..63T, 2012ApJ...745....9Y} of the neighboring hot and warm channels.

\begin{figure}[ht!]
\centering
\plotone{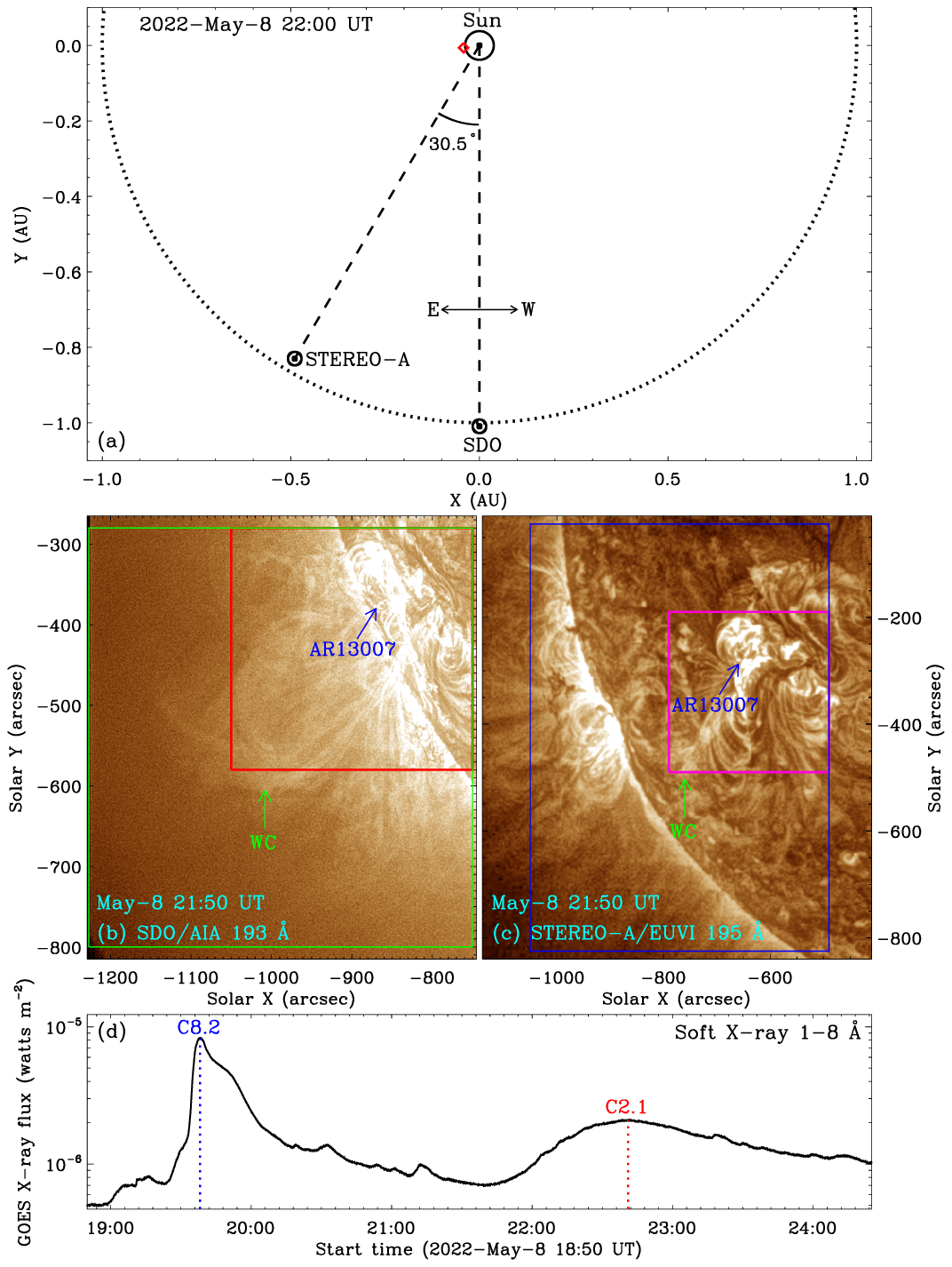}
\caption{General information of the warm channel eruption. 
(a) Positions of the SDO and STEREO-A satellites at 22:00\,UT on 2022 May 8. 
(b) SDO/AIA 193\,\AA~and (c) STEREO-A/EUVI 195\,\AA~images.
(d) GOES-16 1-8\,\AA~soft X-ray flux.
In (a), the dotted semicircle represents the Earth orbit at 1\,AU. 
E and W show the east and west directions in the field of view (FOV) of SDO.
The red diamond marks the general location of the warm channel.
The angle between two satellites is denoted by the number.
In (b), the green and red rectangles separately denote the FOVs of Figures\,\ref{f:warm_channel}(a)-(f) and Figures\,\ref{f:hot_channel}(a)-(g) and \ref{f:flare_cme}(a)-(b).
In (c), the blue and pink rectangles indicate the FOVs of Figures\,\ref{f:warm_channel}(g)-(h) and Figures\,\ref{f:hot_channel}(h)-(i) and \ref{f:flare_cme}(c)-(d), respectively. 
WC in (b) and (c) represents the warm channel.
The blue and red vertical dotted lines in (d) mark the peaks of two flares associated with the eruptions of the hot and warm channels, respectively.
See Section\,\ref{sec:res} for details.
\label{f:general_information}}
\end{figure}

\begin{figure}[ht!]
\centering
\includegraphics[width=1.\textwidth]{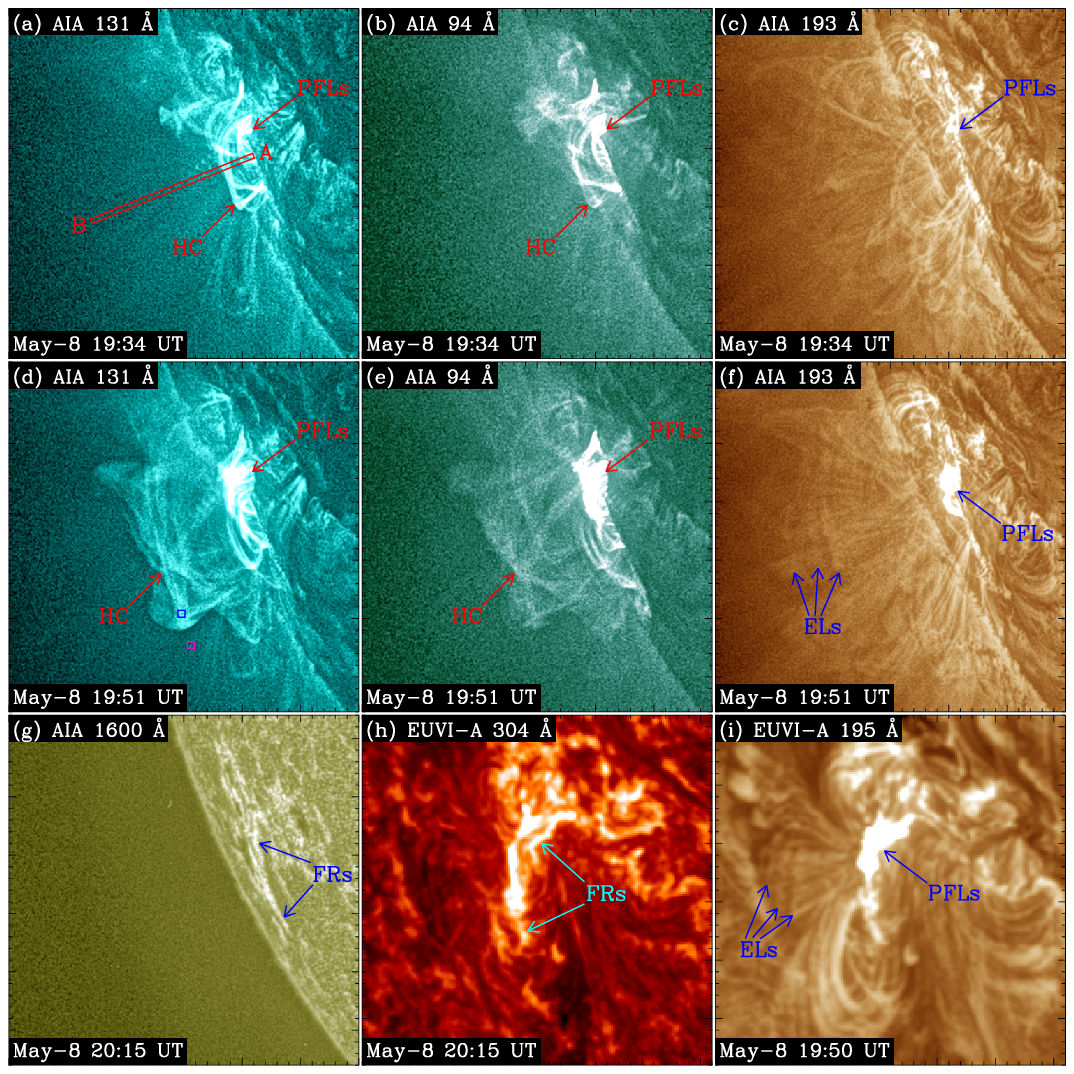}
\caption{Failed eruption of the hot channel before the warm channel eruption. 
(a), (d) AIA 131, (b), (e) 94, (c), (f) 193 and (g) 1600\,\AA, and (h) EUVI-A 304 and (i) 195\,\AA~images.
The red rectangle AB in (a) shows the position for the time slice in Figure\,\ref{f:measurements}(a).
The blue and pink rectangles in (d) enclose the hot channel region for the DEM curve in Figure\,\ref{f:measurements}(c) and the location where the background emission is computed, respectively.
HC, PFLs, ELs and FRs separately represent the hot channel, post-flare loops, expanding loops and flare ribbons.
An animation of the unannotated AIA images (panels (a)-(b)) is available.
It covers 7\,hr starting at 19:00 UT on May 8, with a time cadence of 1 minute.
The FOVs of (a)-(g) and (h)-(i) are indicated by the red and pink rectangles in Figures\,\ref{f:general_information}(b) and (c), respectively. 
See Section\,\ref{sec:hce} for details.
(An animation of this figure is available.)
\label{f:hot_channel}}
\end{figure}

\begin{figure}[ht!]
\plotone{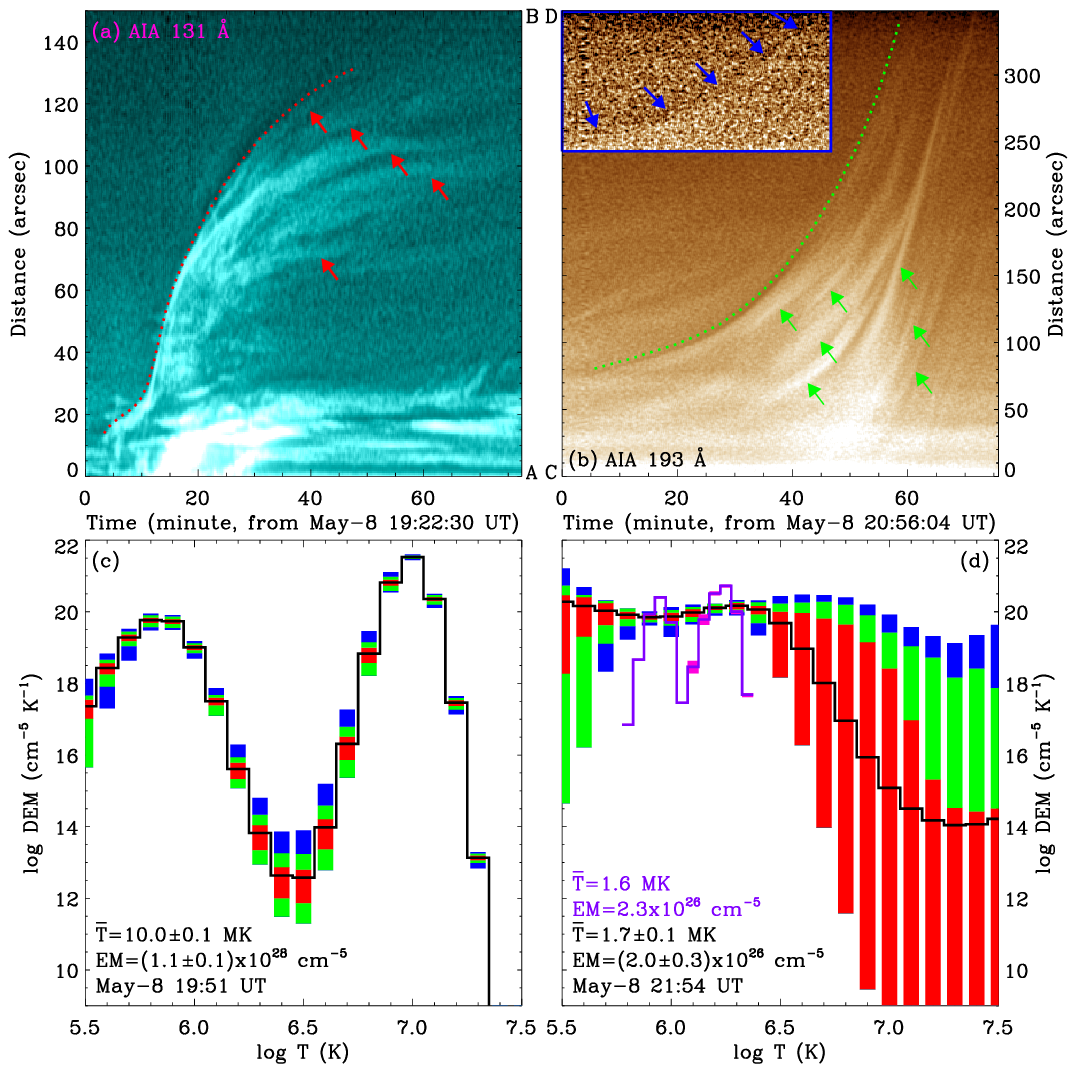}
\centering
\caption{Evolution of the hot and warm channel eruptions. 
(a), (b) Time slices of AIA 131 and 193\,\AA~images along the directions AB and CD in the red rectangles in Figures\,\ref{f:hot_channel}(a) and \ref{f:warm_channel}(b), respectively.
(c), (d) DEM curves separately for the hot and warm channel regions enclosed by the blue rectangles in Figures\,\ref{f:hot_channel}(d) and \ref{f:warm_channel}(e).
In (a) and (b), the red and green dotted lines outline the hot and warm channel eruptions, respectively.
The red and green solid arrows separately mark the hot and warm channels.
In (b), the signal in the blue rectangle is enhanced, and the blue solid arrows denote the compression front.
In (c) and (d), the black curves are the best-fit DEM distributions, and the red, green and blue rectangles represent the regions containing 50\%, 51-80\% and 81-95\% of the Monte Carlo solutions, respectively. 
In (d), the purple curve is the best-fit DEM distribution, and the pink rectangles represent the Monte Carlo solution.
See Sections\,\ref{sec:hce} and \ref{sec:wce} for details.
\label{f:measurements}}
\end{figure}

\begin{figure}[ht!]
\plotone{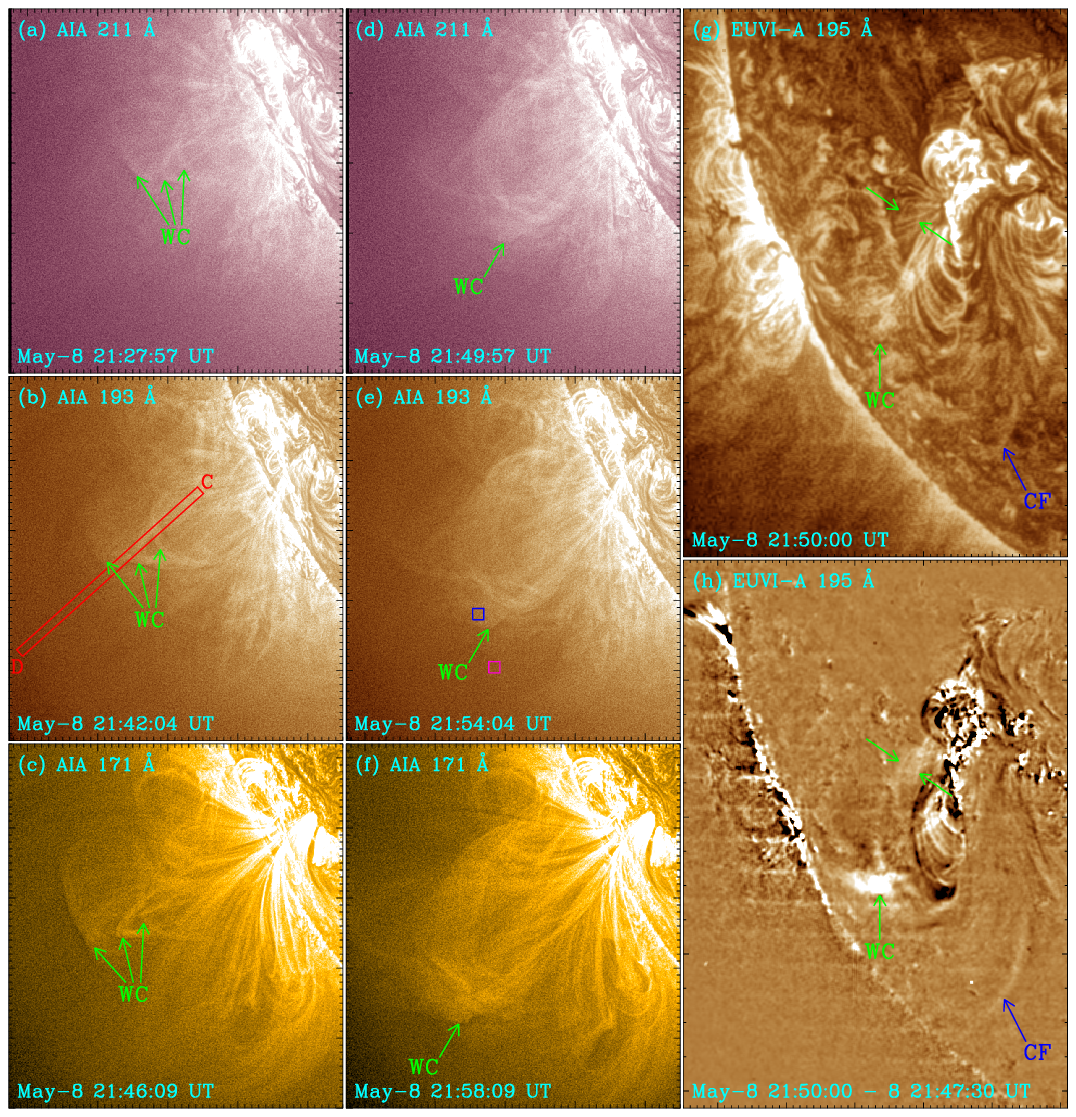}
\centering
\caption{The warm channel eruption.
(a), (d) AIA 211, (b), (e) 193 and (c), (f) 171\,\AA~images, and (g) EUVI-A 195\,\AA~original and (h) running-difference images. 
The red rectangle CD in (b) shows the position for the time slice in Figure\,\ref{f:measurements}(b).
The blue and pink rectangles in (e) separately enclose the warm channel region for the DME curve in Figure\,\ref{f:measurements}(d) and the location where the background emission is computed.
In (g)-(h), the green arrows mark the warm channel, and the CF denotes the compression front.
An animation of the unannotated AIA and EUVI-A (panels (d)-(h)) images is available.
It covers 7\,hr starting at 19:00 UT on May 8, with time cadences of 1 and 2.5 minutes, respectively.
The FOVs of (a)-(f) and (g)-(h) are denoted separately by the green and blue rectangles in Figures\,\ref{f:general_information}(b) and (c).
See Section\,\ref{sec:wce} for details.
(An animation of this figure is available.)
\label{f:warm_channel}}
\end{figure}

\begin{figure}[ht!]
\centering
\includegraphics[width=0.88\textwidth]{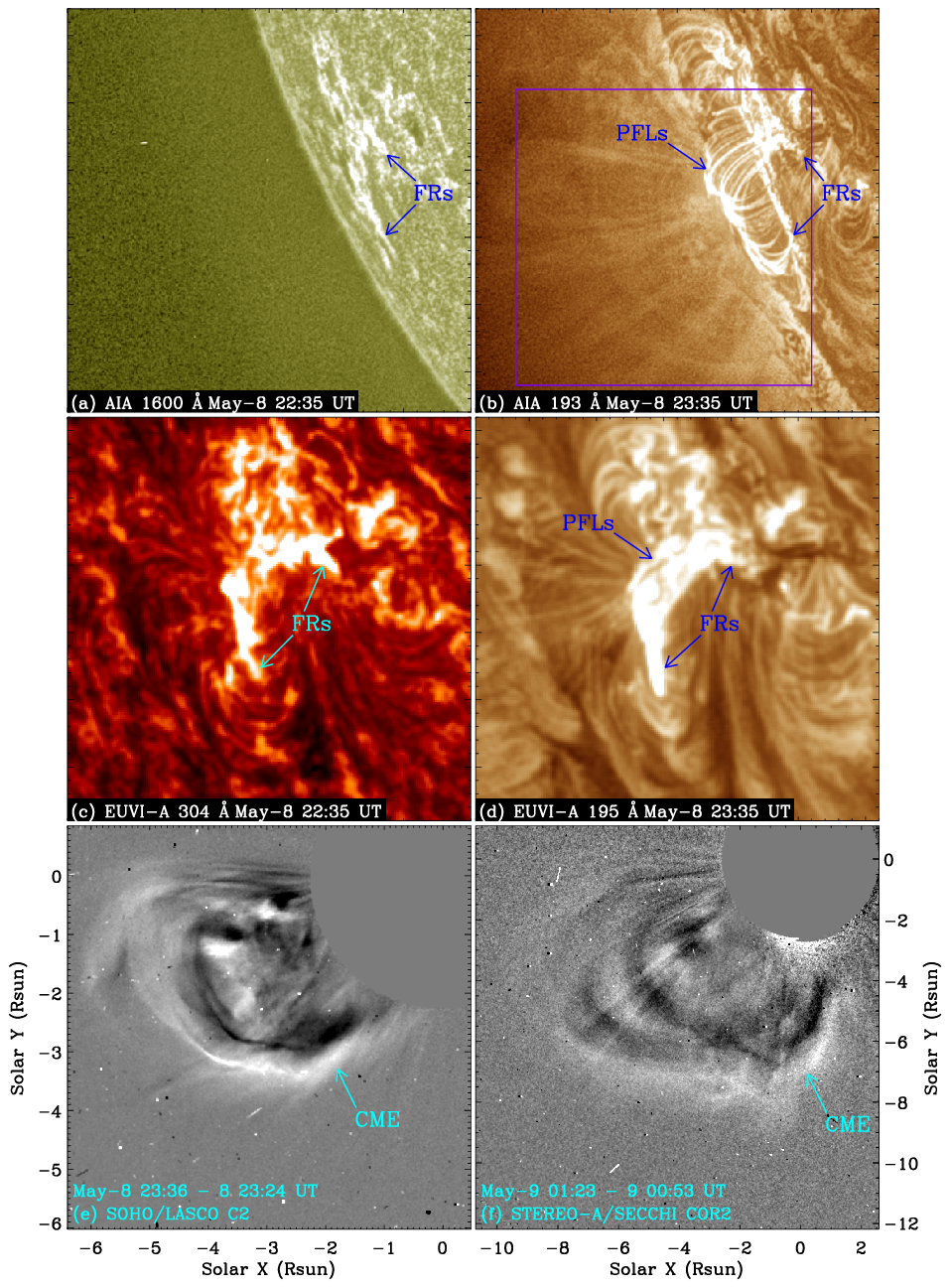}
\caption{The flare and CME associated with the warm channel eruption.
(a) AIA 1600 and (b) 193\,\AA, and (c) EUVI-A 304 and (d) 195\,\AA~images.
(e) SOHO/LASCO C2 and (f) STEREO-A/SECCHI COR2 running-difference white-light coronagraphs.
The purple rectangle in (b) shows the FOV of Figure\,\ref{f:reconnection}.
An animation of the unannotated SOHO and STEREO-A white-light coronagraphs (panels (e)-(f)) is available.
It covers $\sim$4.5\,hr starting at 22:12 UT on May 8, with time cadences of $\sim$12 and $\sim$15 minutes, respectively.
The FOVs of (a)-(b) and (c)-(d) are separately denoted by the red and pink rectangles in Figures\,\ref{f:general_information}(b) and (c).
See Section\,\ref{sec:wce} for details.
(An animation of this figure is available.)
\label{f:flare_cme}}
\end{figure}

\begin{figure}[ht!]
\centering
\includegraphics[width=0.88\textwidth]{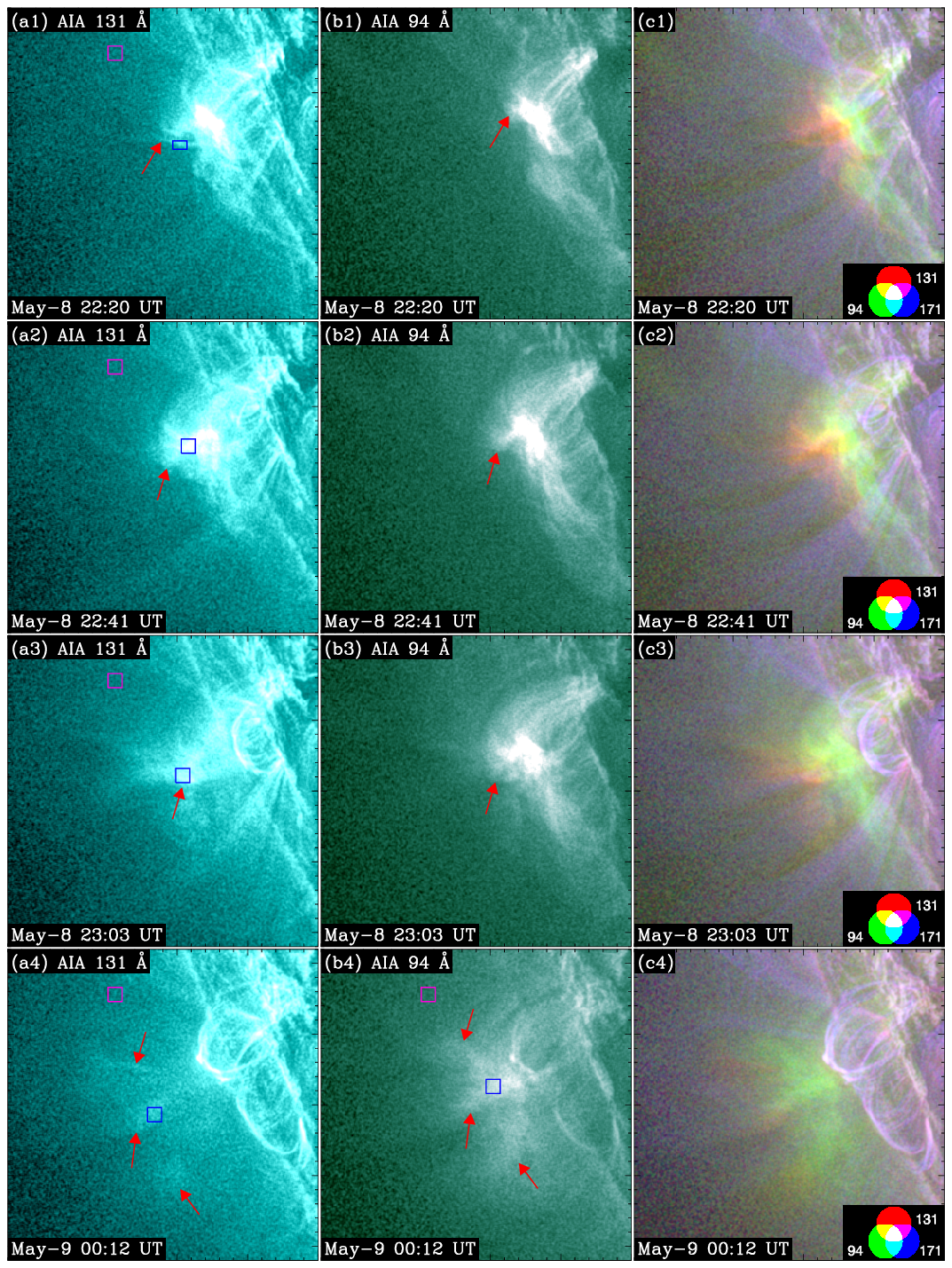}
\caption{Magnetic reconnection above the post-flare loops.
(a1)-(a4) AIA 131 and (b1)-(b4) 94\,\AA~images, and (c1)-(c4) composites of AIA 131, 94 and 171\,\AA~images.
The red solid arrows in (a1)-(b4) mark the reconnection region.
The blue rectangles in (a1)-(a4) enclose the regions for the DEM curves in Figures\,\ref{f:reconnection_evolution}(a)-(d), respectively.
The blue rectangle in (b4) denotes the region for the temporal evolution of the temperature and EM in Figure\,\ref{f:reconnection_evolution}(e).
The pink rectangles in (a1)-(a4) and (b4) show the locations where the background emissions are computed.
An animation of the unannotated AIA images (panels (a1)-(c1)) is available.
It covers 4.5\,hr starting at 21:30 UT on May 8, with a time cadence of 1 minute.
The FOV is indicated by the purple rectangle in Figure\,\ref{f:flare_cme}(b).
See Section\,\ref{sec:wce} for details.
(An animation of this figure is available.)
\label{f:reconnection}}
\end{figure}

\begin{figure}[ht!]
\centering
\includegraphics[width=0.92\textwidth]{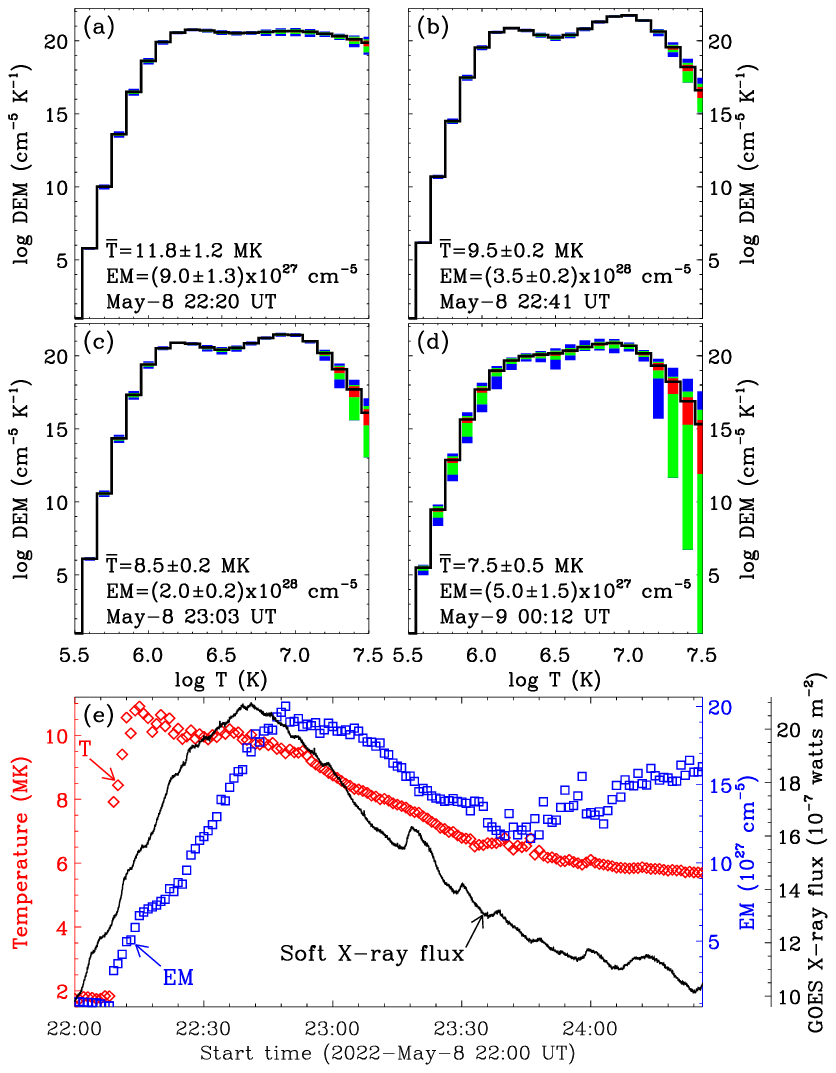}
\caption{Evolution of reconnection above the post-flare loops.
(a)-(d) DEM curves for the reconnection regions separately enclosed by the blue rectangles in Figures\,\ref{f:reconnection}(a1)-(a4).
(e) Temporal evolution of the temperature (red diamonds) and EM (blue squares) of the reconnection region enclosed by the blue rectangle in Figure\,\ref{f:reconnection}(b4).
In (a)-(d), the black curves are the best-fit DEM distributions, and the red, green and blue rectangles represent the regions containing 50\%, 51-80\% and 81-95\% of the Monte Carlo solutions, respectively.
The black curve in (e) shows the GOES-16 1-8\,\AA~soft X-ray flux; see also in Figure\,\ref{f:general_information}(d).
See Section\,\ref{sec:wce} for details.
\label{f:reconnection_evolution}}
\end{figure}

The reconnection region above the post-flare loops during the flare is identified.
Comparing with the previous C8.2 flare associated with the failed eruption of the hot channel, the C2.1 flare associated with the successful eruption of the warm channel lasted for a much longer time; see Figure\,\ref{f:general_information}(d).
This may be the reason why the reconnection region is detected during the smaller, rather than the larger, flare.
The temperature and EM of the reconnection region range in 5.7-11\,MK and 3.1$\times$10$^{27}$-2.0$\times$10$^{28}$\,cm$^{-5}$; see Figure\,\ref{f:reconnection_evolution}(e), consistent with those in \citet{2014ApJ...786...73S} and \citet{2016ApJ...829L..33L}.
The density of the reconnection region, i.e., (2.1$\pm$0.6)$\times$10$^{9}$\,cm$^{-3}$; see Table\,\ref{tab:reconnection}, is, however, smaller than those in \citet{2016NatPh..12..847L, 2016ApJ...829L..33L}.
Both the temperature and EM increased initially, reached the peaks, and then decreased slowly, similar to the evolutionary trend of the soft X-ray flux; see Figure\,\ref{f:reconnection_evolution}(e).
The EM, however, reached the peak $\sim$34 (8) minutes later than the temperature (soft X-ray flux), identical to the results in \citet{2014ApJ...786...73S} and \citet{2016ApJ...829L..33L}.
The difference between the temporal evolution of the temperature and the soft X-ray flux in, e.g., the start and peak times; see Figure\,\ref{f:reconnection_evolution}(e), may be due to the different regions that are chosen separately for calculating the DEM (the small region above the post-flare loops) and the soft X-ray flux (the full solar disk).
Furthermore, material, falling along the reconnection region toward the solar surface, is detected; see Section\,\ref{sec:wce}.
However, no plasmoid is identified in the reconnection region, as the reconnection region is observed face-on, rather than edge-on, from the SDO, consistent with \citet{2016ApJ...829L..33L}.

\clearpage
\startlongtable
\begin{deluxetable}{cccccccccc}
\tablecaption{Information of the hot and warm channel eruptions \label{tab:eruptions}}
\tablehead{
\colhead{MFR} & \colhead{Rising} & \colhead{Erupting} & \colhead{Accel-} & \colhead{Temper-} & \colhead{EM} & \colhead{Width} & \colhead{Number} & \colhead{Associated} & \colhead{Associated} \\
\colhead{} & \colhead{speed} & \colhead{speed} & \colhead{eration} & \colhead{ature} & \colhead{} & \colhead{} & \colhead{density} & \colhead{flare} & \colhead{CME speed} \\
\colhead{ } & \colhead{(km\,s$^{-1}$)} & \colhead{(km\,s$^{-1}$)} & \colhead{(m\,s$^{-2}$)} & \colhead{(MK)} & \colhead{(10$^{26}$\,cm$^{-5}$)} & \colhead{(Mm)} & \colhead{(10$^{8}$\,cm$^{-3}$)} & \colhead{(class)} & \colhead{(km\,s$^{-1}$)}
}
\startdata
hot & 28 & 152 & 221 & 10.0$\pm$0.1 & 110$\pm$10 & 10.8$\pm$0.3 & 32$\pm$2 & C8.2 & no \\
channel & & & -20\tablenotemark{a} &  &  &  &  &  &  \\
\hline
warm & 25 & 206 & 44 & 1.7$\pm$0.1 & 2.0$\pm$0.3 & 29.3$\pm$1.4 & 2.6$\pm$0.3 & C2.1 & 482\tablenotemark{c} \\
channel & & & & 1.6\tablenotemark{b} & 2.3\tablenotemark{b} &  & 2.8$\pm$0.1\tablenotemark{b} & & 593\tablenotemark{d} \\
\enddata
\tablenotetext{a}{Acceleration for the deceleration phase of the hot channel eruption.}
\tablenotetext{b}{Values obtained from the improved sparse inversion code \citep{2015ApJ...807..143C, 2018ApJ...856L..17S}.}
\tablenotetext{c}{Speed obtained from LASCO C2 white-light coronagraphs.}
\tablenotetext{d}{Speed obtained from STEREO-A COR2 white-light coronagraphs.}
\end{deluxetable}

\startlongtable
\begin{deluxetable}{ccccc}
\tablecaption{Information of the magnetic reconnection region during the C2.1 flare associated with the warm channel eruption \label{tab:reconnection}}
\tablehead{
\colhead{Regions} & \colhead{Temper-} & \colhead{EM} & \colhead{Width} & \colhead{Number} \\
\colhead{in} & \colhead{ature} & \colhead{} & \colhead{} & \colhead{density} \\
\colhead{Figure\,\ref{f:reconnection}} & \colhead{(MK)} & \colhead{(10$^{27}$\,cm$^{-5}$)} & \colhead{(Mm)} & \colhead{(10$^{9}$\,cm$^{-3}$)} 
}
\startdata
\ref{f:reconnection}(a1) & 11.8$\pm$1.2 & 9.0$\pm$1.3 & 4.9$\pm$0.4 & 4.3$\pm$0.5 \\
\hline
\ref{f:reconnection}(a2) & 9.5$\pm$0.2 & 35$\pm$2 & 10.9$\pm$0.1 & 5.7$\pm$0.2 \\
\hline
\ref{f:reconnection}(a3) & 8.5$\pm$0.2 & 20$\pm$2 & 15.6$\pm$0.5 & 3.6$\pm$0.2 \\
\hline
\ref{f:reconnection}(a4) & 7.5$\pm$0.5 & 5.0$\pm$1.5 & 12.6$\pm$3.5 & 2.1$\pm$0.6 \\
\enddata
\end{deluxetable}

\clearpage

\clearpage

\clearpage

\clearpage

\clearpage

\clearpage

\clearpage

\acknowledgments

The authors thank the referee for helpful comments that led to improvements in the manuscript, and thank Drs. Chun Xia, Feng Chen, Chaowei Jiang and Qingmin Zhang for discussions. We are indebted to the SDO and STEREO teams for providing the data. 
AIA images are the courtesy of NASA/SDO and the AIA, EVE and HMI science teams. 
This work is supported by the National Key R\&D Programs of China (2022YFF0503002 (2022YFF0503000)), the National Natural Science Foundations of China (12073042, U2031109, 12350004 and 12273061), the Key Research Program of Frontier Sciences (ZDBS-LY-SLH013) and the Strategic Priority Research Program (No. XDB 41000000) of Chinese Academy of Sciences, and Yunnan Academician Workstation of Wang Jingxiu (No. 202005AF150025). 
L.P.C. gratefully acknowledges funding by the European Union (EU). Views and opinions expressed are however those of the author(s) only and do not necessarily reflect those of the EU or the European Research Council (grant agreement No. 101039844). Neither the EU nor the granting authority can be held responsible for them.


\end{document}